# What influences the speed of prototyping? An empirical investigation of twenty software startups

Anh Nguyen Duc[1], Xiaofeng Wang[2], Pekka Abrahamsson[1]

[1]Department of Computer and Information Science (IDI), NTNU
NO-7491 Trondheim, Norway
[2]Free University of Bozen-Bolzano
Piazza Domenicani 3, 39100 Bolzano, Italy
{anhn, pekkaa}@ntnu.no, xiaofeng.wang@unibz.it

**Abstract.** It is essential for startups to quickly experiment business ideas by building tangible prototypes and collecting user feedback on them. As prototyping is an inevitable part of learning for early stage software startups, how fast startups can learn depends on how fast they can prototype. Despite of the importance, there is a lack of research about prototyping in software startups. In this study, we aimed at understanding what are factors influencing different types of prototyping activities. We conducted a multiple case study on twenty European software startups. The results are two folds; firstly we propose a prototype-centric learning model in early stage software startups. Secondly, we identify factors occur as barriers but also facilitators for prototyping in early stage software startups. The factors are grouped into (1) artifacts, (2) team competence, (3) collaboration, (4) customer and (5) process dimensions. To speed up a startup's progress at the early stage, it is important to incorporate the learning objective into a well-defined collaborative approach of prototyping.

## 1 Introduction

With the startup movement, software industry is witnessing a paradigm shift from serving customer requirements to creating customer value. The challenge for software companies is no longer primarily on implementing customer requirements, but rather on finding customer demands and providing a solution that delivers customer value [2]. Addressing uncertainty in both solution and problem domains has often been ad-hoc and based on guesswork, which becomes one of the main reasons for failing startup companies [3]. A demand on systematic approaches to manage the uncertainty has led to an increased research interest on Lean Startup [4], New Product Development (NPD) [5], software startups [6] and continuous experimentation [7].

In a competitive environment such as software industry, time-to-market is becoming more and more critical as a success factor for startup companies. Business ideas under development once revealed can be easily threatened by high speed copycats [9]. Moreover, competitors can also follow an on-going journey of validating product-market fit and arrive faster in the destination. Regardless of company sizes and application domains, the knowledge of influencing factors for a quick learning loop is important for software startups to form best-fit strategy in developing business experimentation [10].

A 'Build-Measure-Learn' loop, as a central concept of the Lean Startup methodology, aims at speeding up the new product development cycle [4]. The central part of the loop is to build a representation of the business, a so-called Minimum viable product (MVP), to collect feedback from customers and to learn from that. Steve Blank emphasizes the goal of MVPs is "*to maximize learning through incremental and iterative engineering*" [2]. In the startup context, developers quickly and iteratively develop a software application to validate business ideas [12]. As such, the study of validated learning can be beneficial from Software Engineering (SE) concepts and practices, such as rapid prototypes and evolutionary prototypes [13, 14, 15]. Consequently, the time-to-release of prototypes is essential to determine the total time in the validated learning loop.

Software startup research is increasingly recognized by researcher's community, with many practical aspects, such as User Experience, Software practices, competences and startup ecosystem [6]. Despite of the importance, there is a lack of research about prototyping in software startups. In a multi-influenced context with funding, human resource and market concerns, it is crucial to understand how the speed of learning can be supported by prototyping activities and what are the influencing factors. In a previous study, we investigated how a prototype is built in software startups [12]. We found that prototyping activities as a core value of startup experimentation needed to be seen as a multifaceted phenomenon [12]. In this work, we are particularly interested in the factors that slow down the learning process and those that speed it up. The research question (RQ) is:

*What factors influence the speed of prototyping in software startups?*

The paper is organized as follows. Firstly, we present the background about business-driven experimentation in software projects, software prototype and a proposal of an analytical model of startup prototyping (Section 2). Then, we describe our research approach and the cases studied (Section 3). After that, the qualitative findings are presented (Section 4). Finally, we reflect on the findings, the threats to validity (Section 5), and draw the conclusion and future work (Section 6).

## 2  Background

### 2.1. Business driven experimentation

From SE perspective, validated learning means the focus on integrating business value in defining software development processes and practices. Even though experiment systems are recognized as beneficial to software projects, there are barriers in adopting them, such as integration of customer feedback, synchronizing vendors in short cycles and lack of reasoning about customer requirements [16, 17]. Bosch et al. [18] advocate for adjusting the Lean startup methodology to accommodate the development of multiple ideas and to integrate them when time for their testing and validation is too long. Bosch suggested using 2-to-4-week experimentation iterations followed by exposing the product to customers in order to collect feedbacks. Fagerholm et al. present a model for continuous experimentation for start up companies [7], in which a key element is the ability to release a prototype with suitable instrumentation, to manage experiment plans, link experiment results with a product roadmap, and to manage a flexible business strategy. Olsson et al. present a Hypothesis Experiment Data-Driven Development model that integrates feature experiments with customer feedback in Agile projects [19]. While these work characterize a process-like approach in developing startups' software products, Paternoster et al. grounded a model from 13 software startups which describes a pattern that software startups often build evolutionary prototypes [20]. This study focuses on how startups are prototyping in reality and the influencing factors of the speed of learning by prototyping.

### 2.2. Prototype and prototyping activities

Brook mentioned "*In software engineering, at least, the concept of rapid prototyping has a name and a recognized value, whereas it does not always have the same status in computer design and in building architecture*" [21]. Prototyping implies a quick and economic approach that serves to achieve understanding of what final products should be [15]. From a technical perspective, prototypes can be differentiated according to its relation to later product development. Throwaway prototypes are used mainly for specification purposes; and they are not used as actual building blocks [15]. They are mostly used in exploratory and experimental prototyping. Evolutionary prototypes provide a basis for a real system, which is evolved out of the prototypes; they are used in evolutionary prototyping but can also be found in experimental prototyping (if it shows that they provide a good basis for a system) [15].

From a business perspective, startups can create a representation of product ideas, a so-called MVP, without actual product implementation. Eric Ries describes a classification of different types of MVPs [4], which are commonly used in the startup communities. For instance, a MVP can be a short animation that explains what a product does and why users should buy it. It can also be a user interface that looks like a real working product, but the actual business process is manually carried out (Wizard of Oz MVP). A concierge MVP is a manual service that consists of exactly the same steps users would go through with the product.

A few research paid attention on improving prototyping activities, such as the speed and effectiveness [28, 29]. Janssen et al. suggested code reuse to speed up writing code to prototype [28]. Grevet et al. described a 6-stage prototyping approach to speed up throw-away prototyping for new social computing systems using existing online systems [29]. In our work, we address the speed of prototyping from a socio-technical perspective, considering prototyping activities under human, market, finance and team factors.

### 2.3. A prototype-centric learning model in software startups

The Build-Measure-Learn loop is a key concept in Lean Startup [4]. The loop is used to manage and to operate software startups in finding a sustainable business model. A key idea is to minimize waste and to focus only on the elements, which will be tested. Lynn et al. describe another cycle, Probe and Learn, that is applicable to manage uncertainties about market, technology and time-to-market [25]. The authors suggest that startups should go to customers with an early version of a product to learn about the market, different applications and segments. Nguyen-Duc et al. propose a hunter-gather double loop to capture the evolution of startup activities from idea to achieving a product market fit [26]. The model visualizes the portion of product development vs. customer development activities across the startup stages. While these studies provide an emphasis on organization and evolution, they are well landed in an abstract space, not straightforward to apply from the SE perspective.

In the SE literature, Gordon et al. propose a rapid prototyping system approach to understand the prototype development of a system [27]. In the model, both low-fidelity and high-fidelity prototypes are essential parts of developing a system [27]. Preliminary product design activities create a throwaway prototype from the problem domain. A series of throwaway low-fidelity prototypes can be created to capture the ideas of what to built. Similarly, high-fidelity prototypes can also be evolved several times before reaching the product launch.

A literature survey of software development shows that startups often build a prototype in an evolutionary fashion and quickly learn from users' feedback [20]. We argue that both throwaway prototypes and evolutionary prototypes are important parts of startups' journey to a launched product. From the Lean startup perspective [4], learning is an input and also an outcome for a prototype. We tailored the double loop model in the previous work [26] by adapting Gordon's system prototyping elements [27] to capture the prototyping processes in the startup context, as shown in Figure 1. The model focus on prototyping as the core concept and compose four loops:

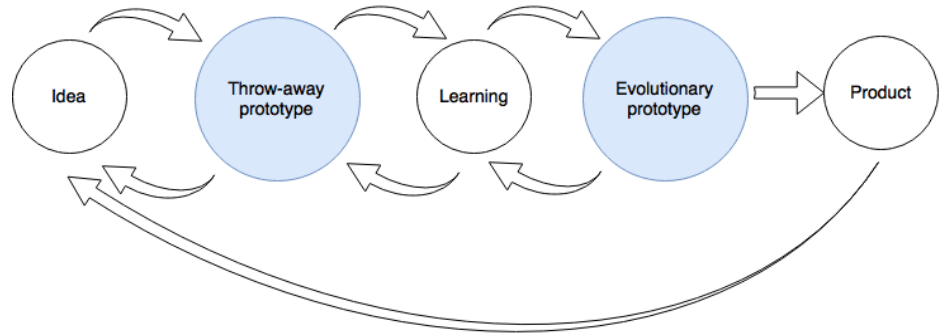

**Figure 1: A prototype-centric learning model in software startups**

- Idea-prototype loop: iterations of refining business idea through throwaway prototyping
- Throwaway prototype loop: iterations of constructing and learning from throwaway prototypes
- Evolutionary prototype loop: iteration of constructing and learning from evolutionary prototypes
- Pivot loop: starting a new cycle from the current product to a pivoted idea

Considering the model as a state-based system, it is possible to travel from a state to any other one. However, the typical flow would happen within two loops. It can also happen that a startup starts the loop from any state, for example, by doing a throwaway prototype before getting to a stated problem. In the scope of this work, we did not go in-depth about how these loops happen in our cases. The work will explore factors that occur during the startup progress and influence throw-away and evolutionary prototyping.

### 3. Research approach

#### 3.1. Multiple case study design

This study is one part of a larger research activity that investigates the role of engineering activities in software startups. The objective is to explore commonalities, challenges and engineering patterns in software startups, from the business idea to a launched product. This study reports the findings from empirical data regarding prototyping activities. We conducted multiple case studies for a robust result in typical software startup population [11]. The unit of analysis is a startup company. We aimed at collecting as many startups as possible for a variety of the sample. As the aim is to reflect the state-of-practice rather than finding a secret recipe of success, we included startups in different stages and with different revenue statuses.

There is often a difficulty in identifying a real startup case among other similar phenomenon, such as freelancers, SMEs or part-time startups. We defined five criteria for our case selection: (1) a startup that operates for at least six months, so their experience can be relevant, (2) a startup that has at least a first running prototype, (3) a startup that has at least an initial customer set, i.e. first customer payments or a group of users, (4) a startup that has an intention to scale their business model, (5) a startup that has software as a main part of business core value.

The process of identifying and collecting data was done in 11 months, from March 2015 to February 2016. Cases were searched from four channels, (1) startups within the professional networks of the authors, (2) startups in the same town with the authors, (3) startups listed in Startup Norway and (4) Crunchbase database. The contact list includes 219 startups from Norway, Finland, Italy, Germany, Netherlands, Singapore, India, China, Pakistan and Vietnam. After sending out invitation emails, we received 41 feedbacks, approximately 18.7% response rate. Excluding startups that are not interested in the research, or startups that do not pass our selection criteria, the final set of cases are 20 startups, aliased as S1 to S20.

### 3.2. Data collection and analysis

Semi-structured individual interviews were used to collect data, since they enable the focus on pre-defined research topics and flexible structures to discover unforeseen information [28]. Methodological triangulation in data collection is also implemented by using evidence from documents and observations (in S01-S05, S09). Business documents, such as business model canvases and business plans were exposed to the research team as a preliminary step prepared for interviews. Observations were useful to understand how prototypes were implemented and used in the working environment.

**Table 1: Startup cases characteristics**

| Code | Product type | Early focus | Later focus | Dev. strategy | No. of prot. | Dev. method. |
|---|---|---|---|---|---|---|
| S01 | Photo marketplace | Feature | | Insource | 2 | Agile |
| S02 | News generator | UX | New feature | Outsource | 4 | Agile |
| S03 | Homemade food market | UX | | Insource | 2 | Adhoc |
| S04 | Construction management | Simple feature | New feature | Outsource | 5 | Distributed Agile |
| S05 | Underwater camera | Feasible technology | | Outsourcing, subcontracting | 7 | Informal Agile |
| S06 | Sale visualization tool | UX | Flexible, scalable | Insource | 3 | Informal Scrum |
| S07 | Location recommendation | Feature, UX | | Insource | 3 | Informal Agile |
| S08 | Ticket platform | Intuitiveness, friendliness | Scalable and new features | Outsource | 2 | Agile |
| S09 | Educational quiz system | User friendliness | Scalable, Stable | Insource | 5 | From adhoc to Distributed Agile |

| S10 | IoT OS platform | Ecosystem | Functionality | Insource | 4 | NO INFO. |
|---|---|---|---|---|---|---|
| S11 | Ticket platform | User friendly, simple | More features, complexity | Insource | 2 | Adhoc |
| S12 | Elearning platform | Feature | | Insource | 3 | Agile |
| S13 | Shipping services | NO INFO. | NO INFO. | Outsource | 3 | NO INFO. |
| S14 | News services | Feature provider | Platform as a service | Insource | 2+ | Continuous development |
| S15 | Smart grid application | NO INFO. | NO INFO. | Insource | NO INFO. | NO INFO. |
| S16 | Secondhand marketplace | innovative feature | Product line | Insource | 3 | NO INFO. |
| S17 | Simulation based training | UX, feature | Flexibility, Scalability | Insource | 2+ | NO INFO. |
| S18 | Open source messenger | Community | Feature | Open source | 4 | Adhoc |
| S19 | Location based alert system | UX | Feature and enhanced UX | Insource | 5 | Agile |
| S20 | Elearning system | User friendliness | Standardization | Insource | 2 | Agile |

*Notation: NO INFO. means missing information*

The interviewees were asked questions about (1) business background (2) idea visualization and prototyping (3) product development (4) challenges and lessons learnt. The stories about startup ideas, prototypes and product development is organized into the schema as described in Figure 1. Most of the interviews were conducted by the first author, with the attendance of a second researcher (the third author or sometimes external researchers in our network). This researcher has a long experience conducting interviews in software companies. Each interview lasted from 55 minutes to 70 minutes and the interviewees were informed about the audio recording and its importance to the study.

We used a thematic analysis – a technique for identifying, analyzing, and reporting standards (or themes) found in qualitative data [22]. We started by reading all interview transcripts and relevant documents, and coded them according to open coding [22]. A set of pre-determined categories were used to guide the coding process, as we have some interests in topics of (1) business original, (2) prototyping practices (3) pivoting (4) testing (5) challenges and (6) key performance indicators (KPIs). We attempted to label all meaningful text segments with appropriate codes. To feed data to this study, we filtered the codes that are related to prototyping, technical implementation, and testing activities prior to product launching. According to Section 2.2, throwaway prototypes were low-fidelity artifacts, such as mockup, wireframe, or simple code. Evolutionary prototypes were perceived as product building blocks, such as heavy code activities, i.e. feasibility testing of functionality, building new feature, etc. The relationship of the factors to the speed of prototyping or production was identified via text about challenges, or text specifying consequence on time-to-market or time to collect user feedback. We noted and reported evidence on prototyping as follows (1) factors that relate to prototyping activities in generals,

(2) factors that slow down the prototyping activities and (3) factors that speed up the prototyping activities.

### 3.2. Case description

The characteristics of our cases are given in Table 1. It is noticeable that a large number of the studied cases deliver peer-to-peer services as marketplaces or platforms (S01, S02, S03, S07, S08, S11, S13, S16, S20). There are also cases that deliver value in Business-to-Business model (B2B) (S04, S06, S10, S12, S15, S17). The cases are dominantly characterized by web-based and mobile-based software product with client-server architecture. We also identified the product focuses in early and later phases of the software startups [23]. Among them, there are some startups with annual revenue of one million euro or more (S06, S09). Regarding the development strategy, interestingly, there are seven cases (35%) that have (parts of) product developed outside company boundary.

The major reported development methodology is Agile, with iterative deliveries and frequent customer feedback: *"… Scrum based development, sprints of two weeks, standup, wrap-up meeting, we like to work in this way."* (S06). In some cases, the company reports a type of informal Agile process: *"… fully informal but truly agile process with working release maintained, … iterative development of functionality and refactoring"* (S05)

One specific question asked to interviewees is how many prototypes have been made before product launching. The answers vary from two to seven prototypes, either throwaway or evolutionary ones, before a launch. In many cases, we considered prototypes as a tangible artifact that is experimented with (potential) users, customers and internal/ external stakeholders.

### 4. Result

Figure 2 describes the influencing elements on throw-away prototyping (detail on Section 4.1) and evolutionary prototyping (detail on Section 4.2). It should be noticed that the direction of impact is not given. Some elements specifically show the positive/ negative influences while other elements remain as general observations.

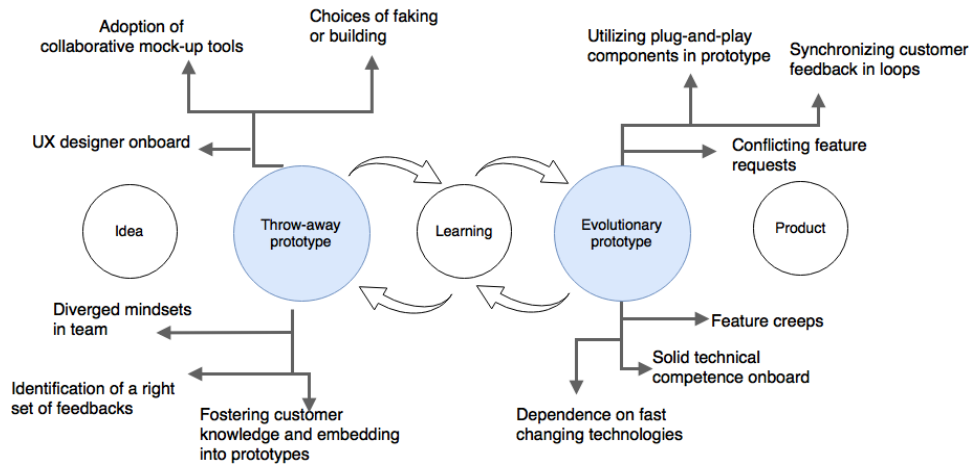

**Figure 2: Factors influencing the prototype-centric learning loops**

### 4.1. Elements influencing throwaway prototyping

#### 4.1.1. Adoption of collaborative mock-up tools

By adopting various tools, i.e. paper sketch, GUI mockups and wireframe tools, startups achieve a fast and an economic prototype without any technical expertise, as described in (S02, S09, S11, S13). In these cases, startups conducted very short iterations, from a few days (S02, S11, S13) to a few weeks (S09), from a product or a service idea to having the first user feedback. In S04, printing GUI layout in papers is reported as a good practice for teamwork, especially improving the customer involvement: "*normally we draw in the piece of paper first and then we make mock-ups... and then the customer joins us on that journey, then we click on the paper, we go to another one …*" (S04). It is also common that startups build mockups by using cloud-based software services. For such an online tool, the teamwork mode is reported as an important feature that facilitates collaborative design efforts among distributed team members (S02).

#### 4.1.2. UX designer onboard

Business side of a startup (often CEOs) is always in a need of expressing and visualizing their ideas into more tangible artifacts. By doing that, sitting next to a designer is highly desirable for CEOs in early stages. In S02, the CEO expresses the need for a close collaboration with a designer in team: "*In this case, I would really like a designer that sits here together with us …*" (S2). The role of a design in mobile application is highlighted in another discussion with S2: "*You might think of user interface as a make-up for a person. But I think UI is the capacity that an app needs to interact with people.*" It happens similarly in S12, when the CEO mentions about the process of designing the graphical part of their prototypes: "*The alternative is to create a specification … and just developing that document and all the process around it is typically very resource intensive. We talk about a future, … we make a* *prototype at a first phase implementation and then we adjust from there based on dialogues in between us*." (S12). For frontend-rich applications, a designer is a champion of the user experience, considering the viewpoint of users and keeping consistency among graphical elements across different platforms.

### 4.1.3. Choices of faking or building

There are often many uncertainties about customers and their expectations in the early stages of startups. Starting with a single-feature prototypes, or other approaches with implementation come always with a risk of wasting effort. It is considered time-saving to start with a clear mind about the throw-away strategy, by focusing on demonstrating business value rather than reusing the technical components (S02). Uncertainty about what to build and how to build often come with quick and dirty experiments without proper architectural designs, appropriate coding practices and documents. In this manner, frequent change of requirements or feature requests could lead to the increase of technical debts in later phase. Experimenting by the development of a runnable prototype was a costly and time-consuming experience in S09. In this way, the value of a prototype should exceed its cost. In S03, the development team has a clear plan for experimenting without "*making the product*" until they get the right product design. S11 applies the concept of "*fake it until you make it*", to simulate a final product without primary quality, both with functionalities and user experience. However, the focus on the speed has also led to the minimum part of viability. In S11, customer demonstration was done in a wizard of oz manner [4], customer interacting with an actual user interface, but business logics and backend functionality were done by manual work. Even though it is inefficient, the approach is easy and fast to build.

### 4.1.4. Collaboration across diverged mindsets

We observed that in most of the cases, the ideas came from the CEOs, who are often business people or serial entrepreneurs. While the decisions about what the products should do come from a business mindset, they are implemented by developers with a technical mindset. In some cases (S01, S04, S05), there are challenges in communicating the product ideas and convincing the developers about the product value. In S04, it took as much time to discuss on the value proposition as to sketch a mockup. Vice versa, the communication of technical difficulties is also a time-consuming task, as mentioned by a developer in S05: "*She [the CEO] is very sharp about business and finance stuffs, but it takes a long discussion to explain her about the importance of having flexible product design …*" (S05). The communication challenge might also happen between startups and customers, when no concrete prototypes are provided: "*We work with a customer organization, learn how they have worked with the current solutions and describe our proposal via the prototype. It is hard for them to realize the benefit without concrete examples…*" (S04). It also appears that a prototype is late released due to the wrong estimation of the CEO, who has no technical background. For example, in S1, the CEO insisted on a customer feedback having a new field in a frontend form, which caused the change of both business logic layer and data table structure.

### 4.1.5. Identification of a right set of feedbacks

Steve Blank emphasizes the importance of early involvement of end users in product development [2]. Particularly, in startups developing products for mass market (or B2C business model), the feedback from the representative users of a market segment is essential. Nevertheless, not all users' input is equally valuable to product development. It was difficult to find the customer feedback that is useful for validating hypotheses in S02: "*I have attended a various types of events like that. To be honest, there are not so many interesting things there …*" (S02). The CEO wandered in town and talked to different people about the product idea. However, the approach is quickly found inefficient, as the users' feedbacks are often shallow. After that, the CEO targeted a group of innovative users from startups and research community and documented many interesting ideas for the product features. The integration of such lead users, "*whose strong needs will become general in a market-place months or years in the future*" [24], appears to be an important factor to accelerate the speed of startup learning. Lead users are also able to contribute via suggestions, testing and feedback, or even participate in the development and co-creation of new products or services, as observed in S14: "*We always do that in a close relation to our actual client stakeholders. Once we decide to narrow it on a new product area, the first thing we do is to get a partnership with a customer so that we can work together on a daily basis as stakeholders and product developers…*" (S14).

### 4.1.6. Fostering customer knowledge and embedding into prototypes

Prototypes can be seen from three different perspectives, function, look-and-feel and role, in which role is the representation of usability of the prototype [2]. In order to maximize lessons learned from a prototype, the vision on how end-users adopt a final product need to be visualized and captured in the prototype. As the actual end users are often not well known in the early phases, the integration of the user's role into the prototype design is a fuzzy task. The time pressure on prototyping makes startups skip a detailed analysis of users' behaviors. It seems that the adoption of customer/ market analysis tools are not so common in our startup sample. In S02, the CEO emphasized the role of mapping tools, such as a customer journey map to describe the customer's experience: *"I have been told by my friends about the tool [a customer journey map]. We used it to describe how customer interact with the system and where could be the gap" (S02).*

### 4.2. Elements influencing evolutionary prototyping

### 4.2.1. Utilizing plug-and-play components in prototype

Utilizing ready-made components, such as Open source software (OSS) libraries and frameworks unlocks the capacity of experimenting functional as well as non-functional features. The adoption of OSS components was mentioned in all of the cases, from using tools (S19), integration of OSS code (S02, S03, S05, S20), to participation in OSS community (S18). The main benefits include reduced development cost and faster time-to-release, which were mentioned by the CTOs of (S19) and (S20): "*…we might not even come to the idea of making it happen if we do *not have OSS as an experiment. Without OSS it would take a lot of time and very costly*" (S19). It is an even more obvious choice in open source type of platforms: "*It is very hard nowadays not to use OSS artifacts, especially when with Android development …*" (S20). It also appears that many advanced technologies were adopted via using OSS: "*A core part of our product includes a machine learning algorithm. We are lucky enough to find ml library in C++, entirely OSS, super cool*" (S02). By taking ready-made components, startups also reduce prototyping time by simplifying architectural aspects to some existing patterns.

### 4.2.2. Synchronizing customer feedback in loops

Communication among team members or between a startup company and its external stakeholders is found as a significant factor delaying an iteration release. Insufficient communication due to misunderstanding, cultural difference, language barrier, lack of supporting tools happens often in outsourcing and remote partnership scenarios (S01, S09): "*Basically, we found some limitations that made it difficult to be efficient in the way to communicate. And since we're teams in different places it's really important that information flow works and also to make sure that all people—don't have to be involved in everything, and be able to group efficiently and create like projects, and store documents, and all these things, and have video-share links, and articles, and all these things*." (S09). The misunderstanding and reworking also happens when customers are distant to developers and the customer feedbacks are not fully perceived. In S13, the CEO and sales people interacted with customers and collected insightful feedback from them. However, the feedback is not communicated efficiently to the development team in other locations. This leads to unnecessary re-work with communication and implementation effort and hence slows down the time to release.

### 4.2.3. Conflicting feature requests

It is a typical situation that evolutionary prototypes are built based on feature requests from the first customers. Gradually, when having more customers, new feature requests might vary from the business direction or even conflict with the previous functionalities. S14 describes how they handled such situation: "*either we solve them by providing them different products or we do ignore parts of the market… We make a very clear statement to what we think the future of journalism is, then we pursue that and the cost of that is neglecting parts of our market*" (S14). Similarly, S15 expresses how their product evolved through different iterations: "*There will always be requirements arriving... Sometimes the new requirements disrupt the old requirements. At the moment, we are working to disrupt the old products*" (S15). Considering what to develop and which features to include adds complexity to future releases. Additionally, requests coming in the middle of the development sprint from large customers might influence the feature priority and delay the release further: "*We're in that situation all the time, it's very difficult to say no because giant customers telling you we need that functionality. If you're going to have us as customers you're going to have to make it, we need it in the contract that you have to make it. We also build it, we built it bigger and bigger*" (S11).

### 4.2.4. Feature creeps

Many startups add new features to fit the prototype to a changing group of early customers. This leads to two possible challenges of satisfying customer demands, so-called (1) feature creep and (2) product portfolio. Feature creep refers to the addition of features to a product in a continuous manner: "*We are adding features all the time. This is not a product that will ever stop evolving. We will always have a strong engineering team to develop the product forward. We are not talking about maintenance here. We are talking about this being the core of the company's competence*" (S13). Startups rarely have a requirement management process to manage product complexity. Consequently, feature creeps are considered harmful to the production and enhancement of core features.

Moreover, this can be an unwanted expansion that requires changes also in the product architecture and even in the strategic direction. In S04, after the first two releases addressing a construction manager's requirements, the third release was developed for a construction operator's demands. Consequently, S04's product scope has grown from a single feature MVP to a supply-chain management system: "*So then we had a small one just for easy communication between users of the building and the maintenance guys… So the second feature was to manage document flow. And the third was to have a 3D model of the building. And all these things here we spent a lot of time and we were building in parallel with different prospects*" (S04).

In a larger scale, the expansion could lead to deriving a product portfolio. Startups face with challenges of keeping both the focus to increase the quality of core delivered values and satisfaction of important customers. While not all good ideas can be turned into features, some ideas are selected to develop further and might become the core value providers for startups.

### 4.2.5. Solid technical competence onboard

In several cases (S09, S01, S03, S06) the technical competence determines the speed of feature releasing. Startups' technical members are required to possess good technical skills and they also need to be productive in an ambiguous development environment: "*We don't hire people basically for them being cheap because we don't have time. Our challenge is time and to be more productive other kind of competing companies … it's much better to have people that can—within a short time, could produce good code*" (S09). It is also important to write code in a clean and structured manner, to be quality-aware in the early phases: "*The back end was pretty good because he had hired my boss at my current company …there was some friction there in how to develop systems between the professional programmer, my boss, and the copy paste programmers. I think that also contributed to it not working.*" (S11). The combination of technical competence and customer understanding is emphasized in another case: "*… It is very hard to find people both good at technology and have a good sense of commercial edge...*" (S08).

### 4.2.6. Dependence on fast changing technologies

Startups often struggle with thriving in a technical uncertainty, whether under market pull or technology push impacts [20]. Due to different reasons, e.g., specific devices, platforms or protocols becoming popular in market, or new technology gaining momentum, there are needs for changing the current product's features to accommodate new technology (S01, S09, S11). In a small scale, for instance, the adoption of new animation effects, a different type of map, etc. leads to an extension of the current or coming iterations. In S02, the development of an IOS application is delayed after the codebase and all dependent libraries were forced to be upgraded to a newer version of Swift. The team took time to resolve all the changes so the next release can be done in Swift 3.0. The technology uncertainty is expected with mobile applications, as stated by the CEO of S11: "…*at the moment we are changing the technology platform. This perhaps has been the biggest challenge we have decided where to stand and make a new platform on development technology... So next generation which will be out in the market place around summer next year will be quite heavily rearranged*. " (S11). In a large scale, the technical change can lead to a change of business directions.

## 5. Discussion

### 5.1. Reflections on the results

We captured what happened during the early phases of the studied twenty software startups. We identified the factors that are found to influence the speed of prototyping across different types of prototypes. They can be grouped into (1) Artifacts, (2) Team competence, (3) Collaboration, (4) Customer and (5) Process dimensions. **Artifacts** include collaborative tools and reusable components. The practices of adopting artifacts are important for saving time of prototyping user interfaces and functionalities. The issue here is to select the suitable tools and components to match the prototyping's purposes. The requirement of **team competence** might vary due to the type of prototyping and the type of products. For instance, UI-rich application would require a designer onboard at the early stage while a good developer in the later stage. **Collaboration**, including efficient communication of visions and tasks among startup teams and interaction with external stakeholders, is important for shorten the learning loops. Besides, **how customers** are involved in the prototyping loops has an impact on the duration of the prototyping. While inappropriate customer feedback delays the learning and creates more prototyping loops, too many requests from customers delay the time-to-release and introduce complexity to product management. Last but not least, prototyping is performed under many uncertainty and dependencies. Defining practices and **processes** to support decision-making under uncertainties would help in prototyping.

### 5.2. Threats to validity

There are several threats to validity worth discussing [1]. One internal threat to validity is the bias in the data collection, as the data might not represent the comprehensive case. This is worth discussing as most of the cases are represented by one interview. In order to mitigate this threat, we selected CTO and CEO as interviewees, who have the best understanding about their startups. We also use other types of data sources, such as documents and observations to increase our understanding about the cases (S01 – S05, S09). The participative observations in S01 and S02 enabled deeper insights that go beyond cross-sectional interviews. A construct validity threat is the possible inadequate descriptions of constructs. We tried at our best to collect contextual information about the startups, from social media and personal contacts. When analyzing data, the coding process of interview transcripts was assisted by the authors' prior knowledge about prototyping and validated learning. This helped to focus on the investigated phenomenon without losing relevant details.

The external validity is normally not addressed by case study research. Our result is grounded on twenty cases, with diversity in company size, application domain, financial model, and growth stage and organization structure, which adds the robustness to our findings. Many themes, such as Section 4.1.1, Section 4.2.1, Section 4.2.5, Section 4.2.6 are observed in more than half of the cases. Our sample is characterized by Norwegian software startups, with a small team and bootstrap financing model. We do not consider other types of startups, for example, internal cooperate startups, venture capital invested startups, and American startups. Hence, the results cannot be directly applied to other contexts, though analytical generalization may be possible in similar contexts.

## 6. Conclusions

To the best of our knowledge, this is the largest multiple case study research about software startups. Grounded on twenty European startups, we adopted an analytical framework to reveal different factors that influence the prototyping activities in early stages of software startups. We found that both throw-away and evolutionary prototypes were influenced by artifacts adoption approach, available team competence, collaboration and customer involvement. Even though there is certain limitation in our case sample, there are still valuable lessons learnt for practitioners. For startups that follow the Lean Startup approach, it is important to align the learning objective with a collaborative and well-defined approach of prototyping. Moreover, startups need to find a systematic approach to integrate relevant external feedback in all phases of prototyping.

This work does not address the evolution of startups according to the learning loops, i.e. what are lessons from idea to throw-away prototype, what are lessons from switching from throw-away prototypes to evolutionary ones. Besides, future work can investigate different types of learning brought by different types of prototypes. This work addressed validated learning through an important angle, which is the speed of prototyping loops. In the future work, we will explore another equally important aspect, which is the quality of learning. Further studies might also identify the effective prototyping and development patterns among software startups.